\documentclass[fleqn,10pt]{wlscirep}

\title{Potential landscape of high dimensional nonlinear stochastic dynamics  with large noise}

\author[1,+]{Ying Tang}
\author[2,+]{Ruoshi Yuan}
\author[3]{Gaowei Wang}
\author[3]{Xiaomei Zhu}
\author[3,*]{Ping Ao}
\affil[1]{Department of Physics and Astronomy, Shanghai Jiao Tong University, Shanghai 200240, China}
\affil[2]{School of Biomedical Engineering, Shanghai Jiao Tong University, Shanghai 200240, China}
\affil[3]{Key Laboratory of Systems Biomedicine Ministry of Education, Shanghai Center for Systems Biomedicine, Shanghai Jiao Tong University, Shanghai 200240, China}

\affil[+]{these authors contributed equally to this work}
\affil[*]{aoping@sjtu.edu.cn}


\begin{abstract}
Quantifying stochastic processes is essential to understand many natural phenomena, particularly in biology, including cell-fate decision in developmental processes as well as genesis and progression of cancers. While various attempts have been made to construct potential landscape in high dimensional systems and to estimate rare transitions, they are practically limited to cases where either noise is small or detailed balance condition holds. A general and practical approach to investigate nonequilibrium systems typically subject to finite or large multiplicative noise and breakdown of detailed balance remains elusive. Here, we formulate a computational framework to address this important problem. The current approach is based on a least action principle to efficiently calculate potential landscapes of systems under arbitrary noise strength and without detailed balance. With the deterministic stability structure preserving A-type stochastic integration, the potential barrier between different (local) stable stables is  directly computable. We demonstrate our approach in a numerically accurate manner through solvable examples. We further apply the method to investigate the role of noise on tumor heterogeneity in a 38 dimensional network model for prostate cancer, and provide a new strategy on controlling cell populations by manipulating noise strength.
\end{abstract}
\begin{document}

\flushbottom
\maketitle
%
%
\thispagestyle{empty}

\section*{Introduction}

\label{sect1}

Studying stochastic dynamics is a central task to understand various natural and experimental phenomena in physics \cite{kramers1940brownian,RevModPhys.62.251}, chemistry  \cite{eyring1935activated}, and biology \cite{mcadams1997stochastic,zhu2004calculating,wilkinson2009stochastic}.
Specifically, stochastic transitions induce current switching in semiconductor \cite{PhysRevLett.109.026801}, reveal population stabilization \cite{PhysRevLett.107.180603} or extinction \cite{PhysRevLett.103.068101},  and provide an integrated picture for genesis and progression of complex diseases such as cancers \cite{wang2014quantitative,zhu2015endogenous}.  Potential landscape \cite{wright1932roles,waddington1957strategy,ao2004potential} can be constructed  for the underlying dynamical systems, and used as a powerful tool to quantify multi-stability and estimate rare transitions. However, a general approach to  achieve this task in practice remains elusive. The first challenge is that real-world systems are intrinsically high dimensional, e.g., gene regulatory network \cite{mcadams1997stochastic,wang2014quantitative}, which makes the simulation \cite{wang2011quantifying} computationally unfeasible in general. In addition, systems may also subject to significant random fluctuations \cite{elowitz2002stochastic,paulsson2004summing,choi2008stochastic} that have functional roles such as driving cell fate decision \cite{zhu2004calculating,lei2015biological,losick2008stochasticity,dar2014screening}. To investigate such noisy effects with robust efficiency in high dimensional systems is another major difficulty.

\begin{figure}[ht]
{\includegraphics[width=1\textwidth]{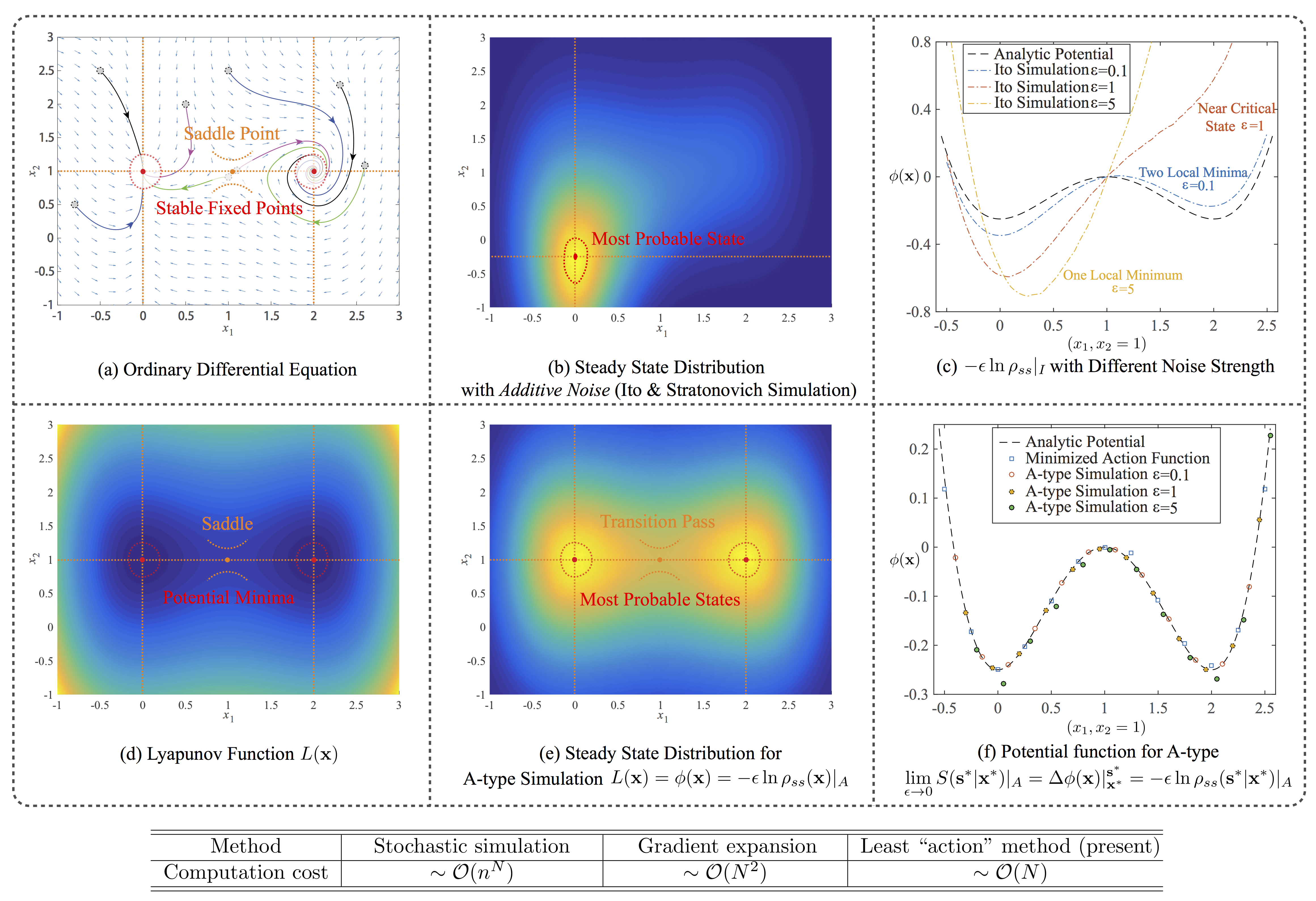}}
\caption{(Color online) The modeling results on the potential (Lyapunov) function and steady state distribution for an illustrative example with 
additive noise. Top panel: the prevailing stochastic simulation leads to a deviated stability structure. Bottom panel: The Lyapunov function, A-type simulation and potential function by least action method consistently reveal stability structure. (a) Vector field showing two stable fixed points and one saddle point. (b) Steady state distribution by Ito's (Stratonovich's) simulation of SDE with $\epsilon=1$, where positions of fixed points from ODE and bistable topology are altered by large noise. (c) Deviations on potential values along the line $x_{2}=1$ between the analytical construction and Ito's simulation of SDE with noise strengths $\epsilon=0.1, 1, 5$. (d) Lyapunov function for ODE. (e) Steady state distribution by A-type simulation of SDE with $\epsilon=1$, where positions of fixed points and bistable topology are consistent with those from ODE and Lyapunov function. (f) Potential values along the line $x_{2}=1$ calculated by least action method, and A-type simulation of SDE with noise strengths $\epsilon=0.1, 1, 5$. Parameters are: $d=a=H=1$. Table: computational costs of the methods on calculating potential barrier between fixed points with respect to system's dimension $N$, where $n$ is the number of mesh points in each dimension for stochastic simulation.}
\label{figure1}
\end{figure}

Previous attempts of simulating the steady state distribution to calculate the potential landscape suffer from exponentially increasing computational cost, and thus direct simulation encounters the curse of dimensionality. The sampling efficiency can be improved when detailed balance condition is employed \cite{PhysRevX.6.011009}, however, this condition breaks down for nonequilibrium systems \cite{PhysRevE.91.042108}. Except simulations, the methods based  on WKB approximation \cite{PhysRevLett.113.078102} or Freidlin-Wentzell theory \cite{freidlin2012random,zhou2012quasi,lv2014constructing,PhysRevX.5.031036} are proposed, but their applications are restricted by the zero noise limit. 
The reason is that they use fixed points of the deterministic model as most probable states for the stochastic process. Nevertheless, such determined fixed point positions can be altered by their implicitly used stochastic integration, typically Ito's or Stratonovich's \cite{gardiner2004handbook}, known as ``noise effects'' \cite{horsthemke1984noise,RevModPhys.79.829,PhysRevLett.111.058102,PhysRevE.90.052121}.
The deviation appears even when noise is additive  (illustrated in FIG.~\ref{figure1}), and becomes dramatic when noise is intensive, which is widely observed in real-world systems \cite{elowitz2002stochastic,paulsson2004summing,choi2008stochastic}. Therefore, on one hand, the high dimensionality defies the use of expensive stochastic simulation; on the other hand, stochastic simulation seems inevitable except when noise is small. This conundrum avoids quantifying stability and stochastic transitions in high dimensional systems with large noise and breakdown of detailed balance.

Towards resolving this problem, we develop a computational framework based on path integral, least action principle  and A-type interpretation \cite{shi2012relation} of the stochastic differential equation (SDE).
We obtain a potential function exactly corresponding to the steady state distribution for SDE and the Lyapunov function \cite{strogatz2014nonlinear} for the deterministic counterpart, an ordinary differential equation (ODE), as exemplified in FIG.~\ref{figure1}(d-f). We further classify two independent causes for the deviation between ODE and SDE when using the prevailing stochastic integrations: 1) the existence of a non-detailed balance part; 2) a variable-dependent diffusion matrix (multiplicative noise). The present method is applicable to both cases, and has two advantages: 1) the computational cost to calculate potential difference is scalable; 2) it is robust under arbitrary noise strength through a general consistency between ODE and SDE, and thus break the small noise restriction.

The present approach can be applied to a wide range of high dimensional stochastic dynamics, and enables us to investigate the role of large noise. It provides an efficient way to calculate probability ratios and transition rates between stable states.
We demonstrate the method in a numerically accurate manner through an example with various noise intensities, and apply it to a 38 dimensional network model for prostate cancer \cite{zhu2015endogenous}.
In particular, the tumor heterogeneity 
\cite{wilkinson2009stochastic,huang2009non,li2015endogenous} is observed controllable by noise intensity. The result may uncover a mechanistic basis for hyperthermia. 

\section*{Results}

\subsection*{Formulation}
\label{sect2}

\subsubsection*{Deviation between ODE and SDE}
ODE and SDE can model a wide range of dynamics \cite{strogatz2014nonlinear,gardiner2004handbook}, for example, chemical reactions \cite{RevModPhys.62.251}, population stabilization \cite{PhysRevLett.107.180603}, and carcinogenesis \cite{zhu2015endogenous,wang2013phage}. ODEs have been successfully used to quantitatively model average behaviors of the stochastic process \cite{barkal1997robustness,wang2014quantitative,gardner2000construction,tyson2003sniffers}, for example, stable states correspond to biological phenotypes \cite{wang2016endogenous}; SDEs, by further adding a vanishing-mean noise term to
capture the source of stochasticity offer a convenient way of examining stability and spontaneous transitions in nonlinear systems \cite{gillespie2000chemical,zhu2004calculating,das2009digital,lei2015biological}. A major advantage of SDE modeling is the extractable dynamical information from the ODE counterpart, which may greatly reduce the computational cost of stochastic modeling.
However, unexpected ``noise effects'' emerge in using SDEs and are reported widely \cite{horsthemke1984noise,RevModPhys.79.829,PhysRevLett.111.058102,PhysRevE.90.052121}. We also demonstrate here that even for a two dimensional example with additive white noise, the dynamical structure can be dramatically altered by noise, e.g., from multi-stable to uni-stable, when applying prevailing simulation methods like Ito's or Stratonovich's \cite{gardiner2004handbook}, as shown in FIG.~\ref{figure1}(a-c). 

To study multi-stability and stochastic transitions, the concept of landscape originally proposed by Wright \cite{wright1932roles} and Waddington \cite{waddington1957strategy} has been developed as a quantitative tool \cite{ao2004potential,wang2011quantifying,freidlin2012random,zhou2012quasi} and are widely used \cite{zhu2004calculating,PhysRevLett.113.078102,lv2014constructing,PhysRevX.5.031036}. Valleys in the landscape correspond to the locally most probable states in the steady state distribution of the stochastic process, and can be recognized as different biological phenotypes \cite{ao2009global}, for instance.
The height of the potential barrier from a valley to another along the landscape corresponds to the transition rate. For a given SDE model, as the unexpected ``noise effects'' when using prevailing stochastic integrations generally deviate valleys from stable states of the ODE counterpart and alter the dynamical structure, one cannot use information obtained by ODE, and has to simulate the steady state distribution to identify the valleys' positions and figure out potential barrier in the landscape, introducing intricacies in applications.

There are two independent causes for the deviation between ODE and SDE with using the prevailing stochastic simulations: 1) the existence of a non-detailed balance part; 2) multiplicative noise. Both causes are related to the freedom of choosing a stochastic interpretation for SDE. With the presence of multiplicative noise, it is known that conventional stochastic integrations like Ito's and Stratonovich's lead to distinct modeling results \cite{gardiner2004handbook}. However, the deviation by non-detailed balance has not been noticed before, and it can occur even for systems with additive noise, where Ito's and Stratonovich's show no difference. Therefore, the prevailing stochastic simulations could not achieve the consistency for systems without detailed balance.
More details are given in Sect.~I of Supplemental Information.

Regarding the issues raised above, one may ask the following question: Is there a possibility to eliminate such unexpected ``noise effects'' and establish a general consistency between ODE and SDE modeling even under large fluctuations? If possible, dynamical information from the ODE counterpart can be inherited by the SDE modeling, such that the valleys' positions in the landscape of SDE is obtainable by calculating stable states of ODE. The consistency is particularly necessary in a scenario where an ODE model is properly constructed and quantitatively correspond to experimental data on average, but are invalidated by the usual way of simulating SDE in reconstituting the original stochastic process and vice versa.

\subsubsection*{Bridging ODE and SDE}
We provide a background on the framework that bridges ODE and SDE:
\begin{align}
\label{ODE}\dot{\textbf{x}}&=\textbf{f}(\textbf{x}),\\
\label{Langevin0}\dot{\textbf{x}}&=\textbf{f}(\textbf{x})+G(\textbf{x})\zeta(t),
\end{align}
where the $N$-dimensional vector $\textbf{x}$ denotes state variables, and $\dot{\textbf{x}}$ represents its time evolution. The deterministic part is $\textbf{f}(\textbf{x})$, and the $M$-dimensional Gaussian white noise $\zeta(t)$ has $\langle\zeta(t)\rangle=0$, $\langle\zeta(t)\zeta^{\tau}(t^{'})\rangle=2\epsilon\delta(t-t^{'})$, where $\epsilon$ is the noise strength playing the role of temperature, the superscript $\tau$ denotes transpose, $\delta(t-t^{'})$ is the Dirac delta function, and $\langle\cdots\rangle$ represents noise average. Here, $G(\textbf{x})G^{\tau}(\textbf{x})=D(\textbf{x})$ defines the symmetric positive definite diffusion matrix $D(\textbf{x})$. The multiplicative noise $G(\textbf{x})\zeta(t)$ models that system state can in turn regulate noise by feedback, or
inhomogeneity of the noisy environment. The results in Sect.~\ref{sect2+} are valid for multiplicative noise. The stochastic integration \cite{gardiner2004handbook} for Eq.~\eqref{Langevin0} is to be specified below.

It is challenging to generally construct Lyapunov function \cite{strogatz2014nonlinear} for ODEs, because $\textbf{f}$ is typically nonlinear and cannot be written directly as the gradient of a potential function $U(\textbf{x})$: $\textbf{f}\neq-\nabla_{\textbf{x}}U$. Even so, a decomposed dynamics equivalent to Eq.~\eqref{Langevin0} was discovered \cite{ao2004potential}, and Eqs.~\eqref{ODE} and \eqref{Langevin0} can be coherently decomposed as:
\begin{align}
\label{ODE decomposed}\textbf{f}(\textbf{x})&=-[D(\textbf{x})+Q(\textbf{x})]\nabla L(\textbf{x}),\\
\label{Langevin_decomposed}\dot{\textbf{x}}&=-[D(\textbf{x})+Q(\textbf{x})]\nabla\phi(\textbf{x})+G(\textbf{x})\ast\zeta(t),
\end{align}
where the matrix $Q(\textbf{x})$ is anti-symmetric with $\nabla\phi^{\tau}Q\nabla\phi=0$, and the asterisk means A-type stochastic integration \cite{shi2012relation}. A Lyapunov function $L(\textbf{x})$ and a potential function $\phi(\textbf{x})$ are constructed, and $L(\textbf{x})$ satisfies $dL/dt\leq 0$ for any trajectory of Eq.~\eqref{ODE} \cite{PhysRevE.87.012708,PhysRevE.87.062109,Ma2014Potential}. For Eq.~\eqref{Langevin_decomposed}, by solving the corresponding Fokker-Planck equation (FPE):
\begin{align}
\label{FPE}
\partial_{t}\rho(\textbf{x},t)=\nabla^{\tau}_{\textbf{x}}[D(\textbf{x})+Q(\textbf{x})][\nabla_{\textbf{x}}\phi(\textbf{x})+\epsilon\nabla_{\textbf{x}}]\rho(\textbf{x},t),
\end{align}
which is obtained from the zero-mass limit on a $2N$-dimensional Klein-Kramers equation \cite{yuan2012beyond}, the steady state obeys Boltzmann-Gibbs distribution  $\rho_{ss}(\textbf{x})=\exp[-\phi(\textbf{x})/\epsilon]$. As the steady state is invariant under transformation $\phi\rightarrow\phi+C$ for any constant $C$, we have chosen $C$ such that the distribution is normalized.

\textit{A-type integration} is defined as the connection between SDE~\eqref{Langevin0} and FPE~\eqref{FPE}, and realized by two explicit limiting procedures: first the usual integration limit and then the zero
mass limit \cite{yuan2012beyond}. Even for systems with additive noise, it is different from the conventional $\alpha$-type stochastic integration \cite{shi2012relation} ($\alpha=0$ is Ito's, $\alpha=1/2$ is Stratonovich's), except that when detailed balance condition holds ($Q=0$) it corresponds to $\alpha=1$. 
An exact transformation from A-type integration to Ito's has been achieved \cite{shi2012relation}. \textit{A-type simulation} can thus be implemented as follows: Eq.~\eqref{Langevin_decomposed} is transformed to be an equivalent SDE under Ito's interpretation:
\begin{align}
\label{Langevin_Ito}
\dot{\textbf{x}}=\textbf{f}(\textbf{x})+\epsilon\Delta\textbf{f}(\textbf{x})+G(\textbf{x})\cdot\zeta(t),
\end{align}
where $\Delta\textbf{f}_{i}(\textbf{x})=\sum_{j}\partial_{x_{j}}[D_{ij}(\textbf{x})+Q_{ij}(\textbf{x})]$, and the dot denotes Ito's integration. Thus, one can simulate Eq.~\eqref{Langevin_Ito} with Ito's scheme to realize A-type simulation for Eq.~\eqref{Langevin0}.

An advantage led by the A-type simulation is that for arbitrary noise strength the sampled steady state distribution of Eq.~\eqref{Langevin0} corresponds to the Lyapunov function and potential function:
\begin{align}
\label{result0}
L(\textbf{x})=\phi(\textbf{x})=-\epsilon\ln\rho_{ss}(\textbf{x})|_{A},
\end{align}
where the subscript A denotes A-type simulation. 

Then, the deterministic part of Eq.~\eqref{Langevin0} has the same decomposition as Eq.~\eqref{ODE}, and positions of fixed points from Eq.~\eqref{ODE} are not changed after added noise. For Ito's or Stratonovich's integrations one cannot recognize fixed points of ODEs from the simulated distribution even for additive noise, as shown in FIG.~\ref{figure1}(a-c). A-type simulation also reserves topology of the landscape for arbitrary noise strength \cite{PhysRevE.87.062109,shi2012relation}.

To implement A-type simulation, the matrix $Q(\textbf{x})$ needs to be solved. The computational cost by the gradient expansion method \cite{ao2004potential} is proportional to the square of systems' dimension. In the following, we provide a new numerical approach to calculate the potential function in Eq.~\eqref{result0}. The computational cost to get the energy barrier is linearly proportional to dimension, rather than square of dimension by gradient expansion method and exponential increasing by stochastic simulation (see analysis in Sect.~\ref{computation cost}), as listed in the table of FIG.~\ref{figure1}.

\subsection*{Efficient calculation on potential difference} 
\label{sect2+}
The essential information for multi-stable systems is the relative stability between stable states, which can be extracted from the potential difference \cite{kramers1940brownian,RevModPhys.62.251}.  
Based on the property that fixed points for ODE and locally most probable states for SDE are identical in our framework, we have the following protocol for a typical situation, where two stable fixed points $\textbf{x}^{*}_{1}$ and $\textbf{x}^{*}_{2}$ are connected by a saddle point $\textbf{s}^{*}$ (Protocol~I):
\begin{enumerate}
  \item Identify positions of the three fixed points from ODE, including two stable fixed points and a connecting saddle point.
  \item 
       Calculate the potential difference between each stable fixed point and the saddle point $\Delta\phi(\textbf{x})|^{\textbf{s}^{*}}_{\textbf{x}^{*}_{1}}$, $\Delta\phi(\textbf{x})|^{\textbf{s}^{*}}_{\textbf{x}^{*}_{2}}$ by the least action method given by Eq.~\eqref{result} below. The potential difference between $\textbf{x}^{*}_{1}$ and $\textbf{x}^{*}_{2}$ is $\Delta\phi(\textbf{x})|^{\textbf{x}^{*}_{2}}_{\textbf{x}^{*}_{1}}=\Delta\phi(\textbf{x})|^{\textbf{s}^{*}}_{\textbf{x}^{*}_{1}}-\Delta\phi(\textbf{x})|^{\textbf{s}^{*}}_{\textbf{x}^{*}_{2}}$.
\end{enumerate}

The path integral formulation for Eq.~\eqref{Langevin0} is applied here to calculate the potential difference. The formulation needs to be consistent with the stochastic integration used \cite{tang2014summing}. For A-type integration, $P(\textbf{s}^{*},T_{2}| \textbf{x}^{*},T_{1})=\int^{\textbf{s}^{*}}_{\textbf{x}^{*}}\mathcal{D}_{A}\textbf{x}\exp\{-S_{T}[\textbf{x}]|_{A}/\epsilon\}$,
where the action $S_{T}[\textbf{x}]|_{A}$ is a function of paths $[\textbf{x}(t)]$ with $\textbf{x}^{*}$ and $\textbf{s}^{*}$ as start point at time $T_{1}$ and end point at $T_{2}$ separately. As we have transformed Eq.~\eqref{Langevin0} to Eq.~\eqref{Langevin_Ito}, it is more convenient to use the equivalent path integral formulation:
\begin{align}
P(\textbf{s}^{*},T_{2}| \textbf{x}^{*},T_{1})=\int^{\textbf{s}^{*}}_{\textbf{x}^{*}}\mathcal{D}_{I}\textbf{x}\exp\{-S_{T}[\textbf{x}]|_{I}/\epsilon\},
\end{align}
where the subscript I means Ito's integration. The measure on paths is defined as: $\int^{\textbf{s}^{*}}_{\textbf{x}^{*}}\mathcal{D}_{I}\textbf{x}\equiv\lim_{K\rightarrow\infty}\prod^{K-1}_{k=1}\int d\textbf{x}^{k}/\sqrt{\det[2\pi\epsilon dtD(\textbf{x}^{k-1})]}$, 
where functions of $\textbf{x}$ take pre-points in each interval. The time is discretized into $K$ segments with $T_{1}=t_{1}<\cdots<t_{k}<\cdots<t_{K}=T_{2}$, each interval being $dt$ and $\textbf{x}^{k}=\textbf{x}(t_{k})$. The action function:
\begin{align}
\label{action_A}
S_{T}[\textbf{x}]|_{A}&=\frac{1}{4}\int_{T_{1}}^{T_{2}}\Big|_{I}dt[\dot{\textbf{x}}-\textbf{f}(\textbf{x})-\epsilon\Delta\textbf{f}(\textbf{x})]^{\tau}D^{-1}(\textbf{x})[\dot{\textbf{x}}-\textbf{f}(\textbf{x})-\epsilon\Delta\textbf{f}(\textbf{x})],
\end{align}
where the integration obeys Ito's rule \cite{gardiner2004handbook}. Note that the present action function is different from that of Freidlin-Wentzell's framework  \cite{freidlin2012random} except when $\epsilon\rightarrow0$. Even for SDE with additive noise, the difference still exists, because $\Delta\textbf{f}_{i}(\textbf{x})=\sum_{j}\partial_{x_{j}}Q_{ij}(\textbf{x})$ can be nonzero for systems without detailed balance.

By using the decomposition in Eq.~\eqref{Langevin_decomposed}, we have:
\begin{align}
\label{derivation}
S_{T}[\textbf{x}]|_{A}&=\frac{1}{4}\int_{T_{1}}^{T_{2}}\Big|_{I}dt(\dot{\textbf{x}}-D\nabla\phi+Q\nabla\phi-\epsilon\Delta\textbf{f})^{\tau}D^{-1}
(\dot{\textbf{x}}-D\nabla\phi+Q\nabla\phi-\epsilon\Delta\textbf{f})
\notag\\&\quad+\int_{T_{1}}^{T_{2}}\Big|_{I}dt\dot{\textbf{x}}^{\tau}\nabla\phi
+\int_{T_{1}}^{T_{2}}\Big|_{I}dt(\nabla\phi)^{\tau}(Q\nabla\phi-\epsilon\Delta\textbf{f})
\notag\\&\geq\Delta\phi|^{\textbf{s}^{*}}_{\textbf{x}^{*}}-\epsilon\int_{T_{1}}^{T_{2}}dtD\nabla^{2}\phi-\epsilon\int_{T_{1}}^{T_{2}}\Big|_{I}dt(\nabla\phi)^{\tau}\Delta\textbf{f},
\end{align}
where we have used $\nabla\phi^{\tau}Q\nabla\phi=0$, 
and Ito's formula \cite{gardiner2004handbook}: $\int_{T_{1}}^{T_{2}}|_{I}dt\dot{\textbf{x}}^{\tau}\nabla\phi=\Delta\phi|^{\textbf{s}^{*}}_{\textbf{x}^{*}}-\epsilon\int_{T_{1}}^{T_{2}}dtD\nabla^{2}\phi$ with $D\nabla^{2}$ denoting $D_{ij}\partial_{x_{i}}\partial_{x_{j}}$. For clarity, we ignore the symbol $(\textbf{x})$ for functions of $\textbf{x}$ in the derivation. The inequality in Eq.~\eqref{derivation} becomes equality for the least action path:
\begin{align}
\label{LAP}
\dot{\textbf{x}}=D(\textbf{x})\nabla\phi(\textbf{x})-Q(\textbf{x})\nabla\phi(\textbf{x})+\epsilon\Delta\textbf{f}(\textbf{x}),
\end{align}
where the trajectory tends to go reversely from $\textbf{s}^{*}$ to $\textbf{x}^{*}$.  
Thus, Eq.~\eqref{action_A} counts the accumulation of noise, whose minimization equals to the ``uphill'' energy $\Delta\phi(\textbf{x})|^{\textbf{s}^{*}}_{\textbf{x}^{*}}$. When the trajectory passes the saddle point, it goes ``downhill'' obeying $\dot{\textbf{x}}=-D(\textbf{x})\nabla\phi(\textbf{x})-Q(\textbf{x})\nabla\phi(\textbf{x})+\epsilon\Delta\textbf{f}(\textbf{x})$, where minimization of Eq.~\eqref{action_A} is zero.

In the limit of $\epsilon\rightarrow0$,
\begin{align}
\label{derivation1}
S_{T}[\textbf{x}]|_{A}&\geq\Delta\phi(\textbf{x})|^{\textbf{s}^{*}}_{\textbf{x}^{*}},
\end{align}
and the least action path follows the time-reversal adjoint dynamics \cite{qian2014zeroth} of Eq.~\eqref{ODE} with decomposition Eq.~\eqref{ODE decomposed}: $
\dot{\textbf{x}}=D(\textbf{x})\nabla\phi(\textbf{x})-Q(\textbf{x})\nabla\phi(\textbf{x})$. Since the dynamics is deterministic, the least action path connecting $\textbf{s}^{*}$ to $\textbf{x}^{*}$ is specified.  Then, minimization of the action function
\begin{align}
\label{minimization}
S(\textbf{s}^{*}|\textbf{x}^{*})|_{A}\doteq\inf_{\{T>0\}}\inf_{\{\textbf{x}(T_{1})=\textbf{x}^{*},\textbf{x}(T_{2})=\textbf{s}^{*}\}}S_{T}[\textbf{x}]|_{A}
\end{align}
equals to the potential difference between $\textbf{s}^{*}$ and $\textbf{x}^{*}$, which is unique if the ideal  least action path following $
\dot{\textbf{x}}=D(\textbf{x})\nabla\phi(\textbf{x})-Q(\textbf{x})\nabla\phi(\textbf{x})$ is found.

With Eq.~\eqref{result0}, we reach a formula to calculate potential barrier:
\begin{align}
\label{result}
\lim_{\epsilon\rightarrow0}S(\textbf{s}^{*}|\textbf{x}^{*})|_{A}=\Delta\phi(\textbf{x})|^{\textbf{s}^{*}}_{\textbf{x}^{*}}=-\epsilon\ln\rho_{ss}(\textbf{s}^{*}|\textbf{x}^{*})|_{A}.
\end{align}
We emphasize that the first equality holds only when $\epsilon\rightarrow0$, but the second equality is for arbitrary noise strength, which is different from that of Freidlin-Wentzell's framework valid for $\epsilon\rightarrow0$ \cite{freidlin2012random}. The significance of Eq.~\eqref{result} is that it enables to obtain potential difference by minimizing the action when $\epsilon\rightarrow0$, and the result exactly corresponds to that of Lyapunov function for ODE and by A-type simulation of SDE with arbitrary noise strength.

\subsection*{Probability ratio and transition rate}
As the steady state obeys the Boltzmann-Gibbs distribution, the probability ratio between stable states is:
\begin{align}
\label{ratio}
\frac{\rho(\textbf{x}^{*}_{2})}{\rho(\textbf{x}^{*}_{1})}=\exp\Big[-\frac{1}{\epsilon}\Delta\phi(\textbf{x})|^{\textbf{x}^{*}_{2}}_{\textbf{x}^{*}_{1}}\Big],
\end{align}
where $\Delta\phi(\textbf{x})|^{\textbf{x}^{*}_{2}}_{\textbf{x}^{*}_{1}}=\Delta\phi(\textbf{x})|^{\textbf{s}^{*}}_{\textbf{x}^{*}_{1}}-\Delta\phi(\textbf{x})|^{\textbf{s}^{*}}_{\textbf{x}^{*}_{2}}$. Equation~\eqref{ratio} is valid under arbitrary noise strength, and thus can show variation of the ratio with different noise intensities. It provides the probability ratios of quantities such as the number of different cell types. Specifically, we apply it to analyze tumor heterogeneity by large noise in Sect.~\ref{sect3}. 

When $\epsilon$ is small compared to the height of the potential barrier $\Delta\phi(\textbf{x})|^{\textbf{s}^{*}}_{\textbf{x}^{*}_{1}}$, the asymptotic transition rate formula \cite{kramers1940brownian,RevModPhys.62.251} from the stable fixed point $\textbf{x}^{*}_{1}$ to $\textbf{x}^{*}_{2}$ is:
\begin{align}
\label{transition rate}
R(\textbf{x}^{*}_{2}|\textbf{x}^{*}_{1})&\propto\exp\Big[-\frac{1}{\epsilon}\Delta\phi(\textbf{x})|^{\textbf{s}^{*}}_{\textbf{x}^{*}_{1}}\Big].
\end{align}
Different from \cite{PhysRevE.48.931}, in our framework the non-detailed balance part does not provide correction terms that explicitly appear in the pre-factor of the rate formula as analyzed in Sect.~IV of Supplemental Information. %

\subsection*{Comparison on computational cost}
\label{computation cost}

Computational costs of methods mentioned above for systems with respect to dimension $N$ are analyzed here. For stochastic simulation of Eq.~\eqref{Langevin_decomposed}, we mesh each dimension into $n$ points, and the computational cost is exponentially proportional to dimension, $cost\sim\mathcal{O}(n^{N})$. This method also has the problem of slow convergence when noise strength is small. For the gradient expansion \cite{ao2004potential}, as in each step a matrix needs to be evaluated, the computational cost is approximately $cost\sim\mathcal{O}(N^{2})$. The least action method here needs a one dimensional path connecting two points, and thus its cost is linearly proportional to dimension, $cost\sim\mathcal{O}(N)$. Therefore, the least action method is efficient in high dimensional systems. We list the results in the table of FIG.~\ref{figure1}.

\subsection*{Protocol to obtain a global landscape}
\label{ProtocolII}

Protocol I can be extended to obtain a global landscape for systems with multiple stable states (Protocol~II):
\begin{enumerate}
  \item Identify positions of all fixed points under consideration from ODE.
  \item Choose a saddle point as reference. Starting from points in small neighborhood of the saddle point, find all stable fixed points reached by simulating ODE. Calculate potential difference between the saddle point and the stable fixed points by Eq.~\eqref{result}.
  \item Repeat step $2$ for all saddle points. Fix relative potential difference between the saddle points if they reach common stable fixed points by step $2$.
  \item For any other points in state space, find the fixed point that it reaches by simulating ODE. Obtain their potential difference by Eq.~\eqref{result}.
\end{enumerate}
The consistency between ODE and SDE enables to utilize information of fixed points and basins of attraction from ODE, which greatly improves the efficiency of our algorithm. As each point in state space reaches a single fixed point, its potential value is uniquely determined if the ideal  least action method is found numerically during minimization procedure, which leads to a global landscape without ambiguity. Specifically, the probability ratio between a fixed and a point within the potential well is calculated by Eq.~\eqref{ratio}, which corresponds to cell-to-cell variability inside the attractor. For dynamical systems with complex attractors such as limit cycle \cite{strogatz2014nonlinear,ge2012landscapes,PhysRevE.87.062109}, the point on the stable limit cycle can be treated similarly as the stable fixed point, and thus our method can be generalized. 

\subsection*{Steps of applying the present method}
As the computational cost to obtain potential differences between states is scalable, our method is applicable to high dimensional systems. The steps to use our method is:
\begin{enumerate}
  \item Identify positions of the fixed points from ODE, and classify them into the set of stable and unstable (saddle) points. 
  \item Calculate the potential difference between fixed points by Protocol~I. Extract probability ratios  and transition rates by Eqs.~\eqref{ratio} and \eqref{transition rate} separately.
  \item Use Protocol~II to get a global landscape if needed.
\end{enumerate}
We study the biological examples in Sect.~\ref{sect3} below and Supplemental Information according to these steps.

\subsection*{Application}
\label{sect3}

Heterogeneity of cell populations is widely observed in biological systems such as cancer \cite{hanahan2011hallmarks}, where different cell phenotypes emerge in tumor tissues \cite{junttila2013influence,li2015endogenous}. It is proposed that an underlying regulation network and quantification on the network dynamics by SDE models can describe various cell types and transitions between them \cite{wilkinson2009stochastic,wang2011quantifying}. Starting from the SDE model, the calculated potential landscape provides an integrated picture to study heterogeneity. Specifically, valleys in the landscape correspond to different cell types, and the potential barrier separating them quantifies the transition rates. This approach of landscape is helpful to understand systematically the effect of perturbations on cell type interconversions.

Here, we investigate whether the variation of noise strength leads to changes of cell types, as noise plays a crucial role in biological processes, for example, it drives the cell fate decision \cite{losick2008stochasticity,dar2014screening}. The previous methods \cite{PhysRevLett.113.078102,freidlin2012random,zhou2012quasi,lv2014constructing,PhysRevX.5.031036} can not be applied to study the function of large noise, because identification on valleys of landscape and calculation on potential barrier by these methods are restricted to the zero noise limit.
Now, we are able to quantify the role of large noise on heterogeneity in high dimensional network dynamics, because the present calculation on landscape is robust under arbitrary noise strength. From our method,
the ratios of cell types can be controlled by manipulating noise strength, 
which allows the cell-to-cell variability under the same gene regulation network.



\subsubsection*{Prostate cancer model}
\begin{figure}[ht]
{\includegraphics[width=1\textwidth]{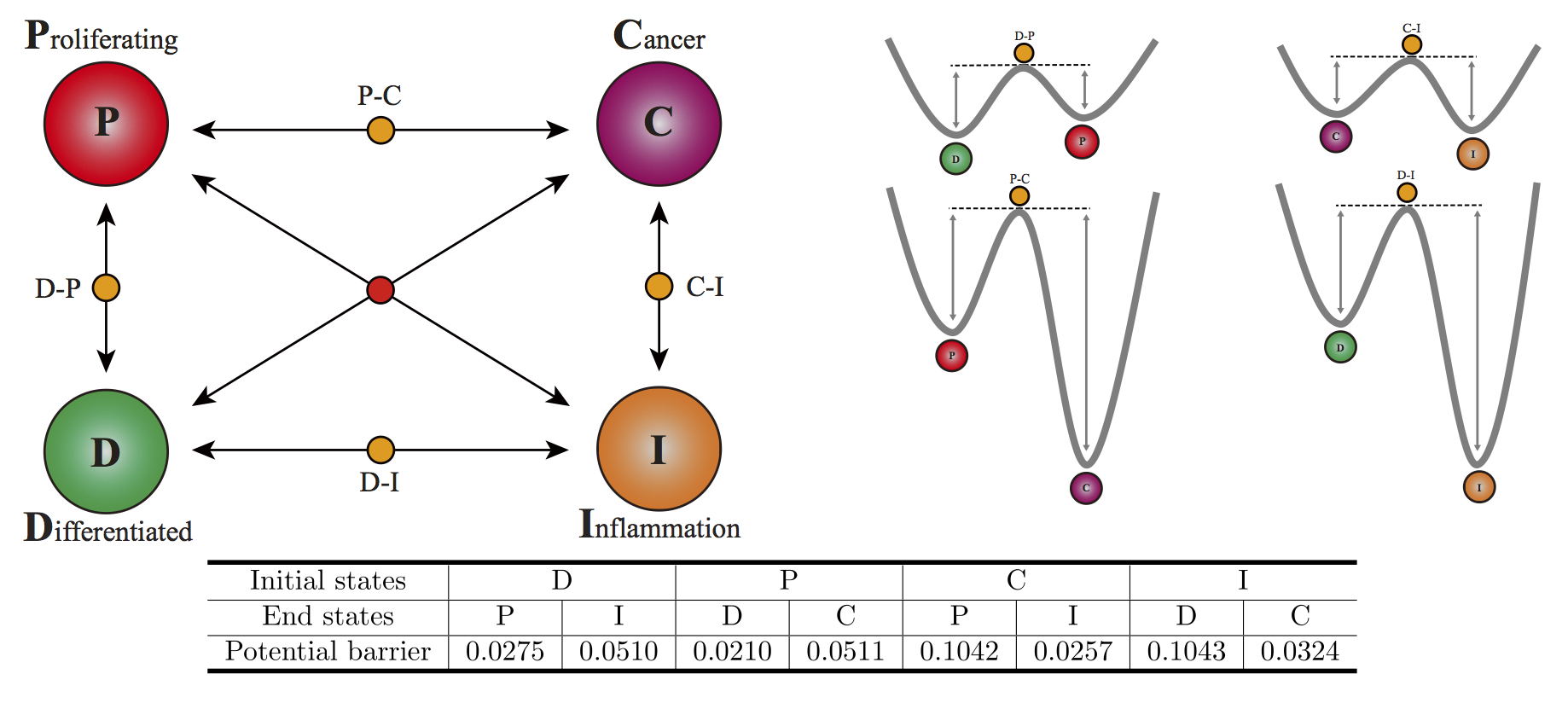}}
\caption{(Color online) Potential barriers calculated by  Protocol~II  in the prostate cancer model \cite{zhu2015endogenous}. Left panel: The chosen four cell types: differentiated (D), proliferating (P), cancer (C), inflammation (I). They are stable fixed points obtained from ODE of the 38 dimensional system. The states D-P, D-I, C-I, P-C are saddle points, and the red fixed point in the middle is unstable. Right panel: the heights of potential barriers between stable fixed points connected by saddles. The lengths of arrows are proportional to barrier heights listed in the table below. Table: potential barriers between stable fixed points are calculated by the least action method.
We set $K=100$, $T=20$, and have checked that larger $K$ and $T$ values lead to convergent results.
 The parameters of the system chosen here are for typical cancer patients \cite{zhu2015endogenous}, where cancer and inflammation states are more stable. }
\label{figure2}
\end{figure}

As another illustrative example, we demonstrate the effectiveness of our approach by applying it to a network model for prostate cancer \cite{zhu2015endogenous}.
The network dynamics is modeled by a 38 dimensional SDE \cite{zhu2015endogenous}. The ODE counterpart and the parameters are given in Sect.~VII of Supplemental Information. From analysis on the ODE, we know the system has 10 stable fixed points and 16 saddle points. To exemplify the method, we consider four stable fixed points corresponding to various cell types shown in FIG.~\ref{figure2}: differentiated (D), proliferating (P), cancer (C), inflammation (I) with their positions given in \cite{zhu2015endogenous}. For clarity,
we consider additive noise case with choosing $D(\textbf{x})$ as identity matrix in this example. It should be emphasized that the deviation between ODE and SDE appears even with additive noise, because this system does not obey detailed balance condition. As a result, considering additive noise is sufficient to demonstrate the advantage of our method based on A-type integration compared with the prevailing methods based on other stochastic integrations.

\begin{figure}[ht]
{\includegraphics[width=0.8\textwidth]{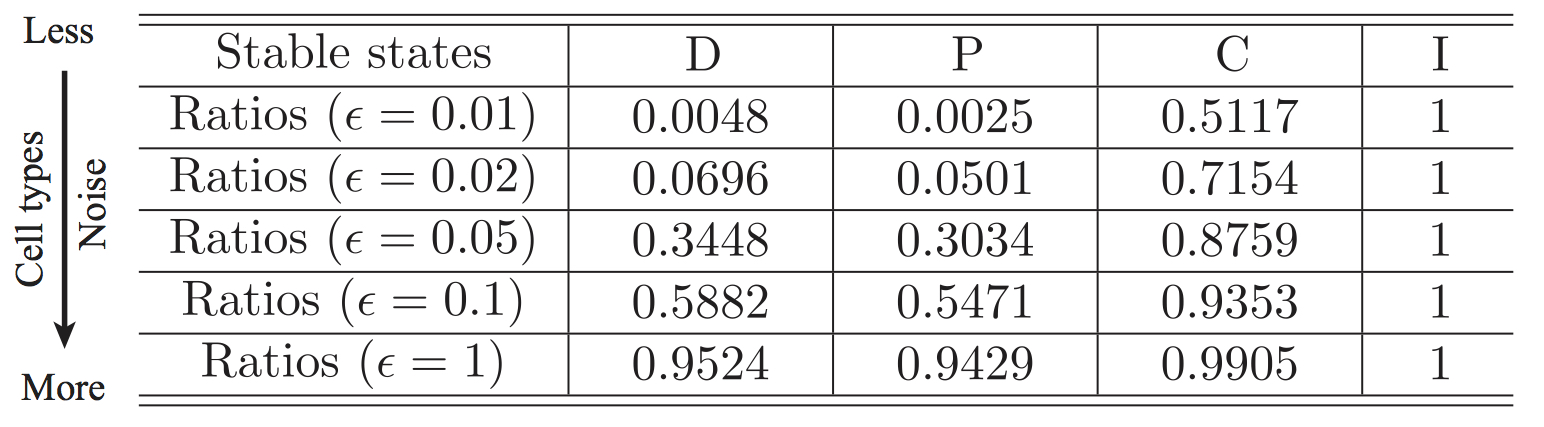}}
\caption{The probability ratios between D, P, C, I states of the prostate cancer model. The ratios are calculated by Eq.~\eqref{ratio} with heights of potential barriers listed in the table of FIG.~\ref{figure2}, where the values are normalized by the I state. We consider noise intensities: $\epsilon=0.01, 0.02, 0.05, 0.1, 1$. The tumor heterogeneity emerges when noise strength becomes large where the four types of cells are almost equally distributed.}
\label{figure3}
\end{figure}

We use the least action method to calculate heights of the potential barriers between stable states, as shown in FIG.~\ref{figure2}. We note that in order to have the global landscape, we use the continuous condition to set the potential value of $P$ to $D$ specifically. According to Eq.~\eqref{ratio}, the probability ratios of different cell types can be calculated under various noise strengths. We list the ratios between the states of D, P, C, I in FIG.~\ref{figure3}, which shows qualitative different results with changing noise. When noise is small, e.g., $\epsilon=0.01$, most of the cells belong to the cancer and the inflammation states, whereas little are in states of differentiated and proliferating. When noise becomes large such as $\epsilon=1$, various cell types are almost equal in number. This demonstrates the emergence of tumor heterogeneity with respect to increasing noise strength.

We elucidate more on the application of controlling the ratios of cell types through varying noise intensity, which can be implemented by tuning temperature. 
First, it was demonstrated that an therapy with combination of hyperthermia and other treatments, such as immunotherapy and radiotherapy, can improve the efficiency of cancer cure \cite{wust2002hyperthermia}. In those cases, temperature plays the role of enhancer to switch on and off the effectiveness of other therapies. Specifically, drug cytotoxicity triggered by temperature variation leads to death of tumor cells, 
and therefore combination of hyperthermia and chemotherapy is regarded as an effective treatment of cancer \cite{issels2008hyperthermia}. Second, as different levels of heating were found to bring distinct modulatory effect on tumor targets, our method valid for arbitrary noise strength may be applied to study sensitivity of thermal treatment regulated by temperature, which will provide new designs on clinical trials. Third, the regional hyperthermia to radiotherapy \cite{hildebrandt2002cellular} shows an improvement on survival rates of cancer patients, because hyperthermia can guide the action of chemotherapy to specific heated tumor region. This can be modeled by multiplicative noise, as exemplified in Sect.~I of Supplemental Information. Therefore, the present approach could form the theoretical basis for hyperthermia that employs effect of temperature in tumor treatment.

With the obtained potential barrier, transition rates of cell type interconversions is given by Eq.~\eqref{transition rate}, which provides a quantitative understanding on cancer genesis. Under the given parameters and noise strength, the result provides a set of predictions: 1) cancer and inflammation states are more stable than proliferating and differentiated states; 2) transitions from cancer state to proliferating state and from inflammation state to differentiated state are difficult than the other way around; 3) transition to cancer state from inflammation state is more frequent than from proliferating state. These suggest that the model describe a cancer patient, and new strategies for medical treatments should be designed to rise the potential energy of cancer and inflammation states. 

\section*{Discussion}

\label{diss}

For a physical system with clearly separated sources for the deterministic force and the stochastic force, stochastic interpretation for SDE (Langevin equation) is chosen by nature. Experiments \cite{lanccon2001drift,volpe2010influence} have shown that a class of systems chooses the anti-Ito's integration \cite{shi2012relation}, corresponding to A-type in one dimension.
For effective models without a clear-cut distinction between deterministic and stochastic components, each stochastic interpretation has its own advantage. For example, Stratonovich's interpretation enables the use of ordinary calculus. The correspondence between ODE and SDE modeling under arbitrary noise strength is a unique property for A-type integration.

In biological problems, noise has a variety of sources \cite{raser2005noise}, 
such as locations of molecules, micro-environmental fluctuations, gene expression noise, and cellular processes like cell growth. 
For complex systems like cancer, noise may come from different sources. SDE model reconstitutes the random fluctuations into a single noise term, which reflects the various sources of noise \cite{suel2006excitable,lei2015biological}.
Therefore, several experimental operations can implement the change of the noise strength discussed here in real biological systems.

Our method can be applied to systems that are modeled by master equation (CME) with discrete dynamical variable \cite{gardiner2004handbook}. 
First, CME can be transformed to be the chemical Langevin equation with continuous variable \cite{gillespie2000chemical}, which can be cast into the form of Eq.~\eqref{Langevin0}. Then, our method is applicable to improve efficiency. The approximation is tolerably accurate when the copy number of variables are large, and it also requires that the dynamical process has a time scale during which multiple reactions occur and the reaction rate does not change dramatically \cite{gillespie2000chemical}.  These conditions are expected to hold for the present  high dimensional cancer dynamics \cite{wang2014quantitative,zhu2015endogenous}, where the proteins usually has high copy numbers.
Second, CME may also be expanded to a FPE and further corresponds to Eq.~\eqref{Langevin_decomposed} with consistent modeling predictions, as demonstrated through an explicit procedure in Sect.~II of Supplemental Information. Third, for systems with low copy numbers, SDE can still provide an appropriate description on the effect of noises \cite{lei2015biological}. Fourth, for stochastic processes on the level of single molecules, such as gene burst process \cite{pedraza2008effects}, CME is a more proper approximation to capture the discrete nature of species \cite{PhysRevLett.103.068101}. Nevertheless, this kind of noise will diminish by accumulation of proteins with long lifetime \cite{eldar2010functional}. 

Mathematically, SDE and CME are two independent modeling methodologies, and are on an equal footing to describe the stochastic dynamics. Both CME and SDE are models with intrinsic discrepancy to the real dynamical process.
From computational side, a whole set of CME to describe the stochastic dynamics in detail is typically high dimensional, and the Gillespie algorithm \cite{gillespie1977exact} to simulate CME is time consuming. Thus, the present method handling SDE valid for arbitrary noise strength is practically useful to investigate high dimensional systems with large fluctuations, particularly when the ODE counterpart is properly constructed and quantitatively correspond to the average experimental data. 

Several other remarks are in order.
First, our calculation on potential difference is applicable to systems both with and without detailed balance condition \cite{ao2004potential}, i.e. $Q=0$ or not. Breakdown of detailed balance inducing a curl flux in the state space affects the least action path, and generally leads to $S(\textbf{x}^{*}_{1}|\textbf{x}^{*}_{2})|_{A}\neq S(\textbf{x}^{*}_{2}|\textbf{x}^{*}_{1})|_{A}$. For such cases, the least action path also differ from the deterministic saddle-node trajectories \cite{freidlin2012random}. Second, there are many  efficient numerical methods to calculate fixed points of ODEs in high dimension \cite{ascher1998computer}, such as Newton iteration method. Third, the present action function has the dimension of energy, and the conventional action in classical physics has the dimension of energy multiplied by time. Fourth, positions for the locally most probable states in Eq.~\eqref{ODE decomposed} is a subset of fixed points' positions for Eq.~\eqref{ODE}, because $\nabla \phi(\textbf{x})=0$ is sufficient  but not necessary to $\textbf{f}(\textbf{x})=-[D(\textbf{x})+Q(\textbf{x})]\nabla \phi(\textbf{x})=0$. Fifth,  the constructed potential function is also useful to extract thermodynamical free energy for non-equilibrium systems \cite{PhysRevE.92.062129,PhysRevE.89.062112}.

We next compare our framework with the previous works. First, authors in \cite{PhysRevLett.113.078102} used a WKB method and reached a Hamiltonian-Jacobi equation $\nabla\phi(\textbf{x})^{\tau}D(\textbf{x})\nabla\phi(\textbf{x})+\nabla\phi(\textbf{x})^{\tau}\textbf{f}(\textbf{x})=0$ in the lowest order of $\epsilon\rightarrow0$. However, computational cost of solving this partial differential equation increases exponentially. This Hamiltonian-Jacobi equation is valid for arbitrary orders of $\epsilon$ in our framework \cite{yuan2012beyond}. Second, our method is different from the previous path integral approach \cite{wang2011quantifying}, where their action function gives an effective potential rather than the exact potential $\phi(\textbf{x})$ constructed consistently in Eq.~\eqref{ODE decomposed} and Eq.~\eqref{Langevin_decomposed}. Third, the unexpected ``noise effects'' in using SDEs have been widely reported \cite{horsthemke1984noise,RevModPhys.79.829,PhysRevLett.111.058102}, and whether the phenomena are produced by the physical effect of the zero-mean noise or by the intricacy of using various stochastic integrations \cite{PhysRevE.90.052121,PhysRevLett.115.240601} (also see the example in Sect.~II of Supplemental Information) is a question without a definite answer. These effects in general defy the use of dynamical information from the ODE counterpart, and our method provide a possibility to reserve useful results by ODE analysis for SDE with arbitrary noise strength. 

We analyze the difference between our framework and Freidlin-Wentzell's \cite{freidlin2012random}, based on which the quasi-potential has been calculated in many systems recently \cite{lv2014constructing,zhou2012quasi,PhysRevX.5.031036}. First, the consistency between action function's form and stochastic integration (classified in table~II of Supplemental Information) has not been considered in Freidlin-Wentzell's framework. Only the present action function of Ito's form is the same as the usual Freidlin-Wentzell's action \cite{freidlin2012random}. Second, our decomposition $\textbf{f}(\textbf{x})=-[D(\textbf{x})+Q(\textbf{x})]\nabla\phi(\textbf{x})$ with $\nabla\phi(\textbf{x})Q(\textbf{x})\nabla\phi(\textbf{x})=0$ is generally different form the usual Freidlin-Wentzell form $\textbf{f}(\textbf{x})=-\nabla U(\textbf{x})+l(\textbf{x})$ with $\nabla U(\textbf{x})\cdot l(\textbf{x})=0$, except when diffusion matrix $D(\textbf{x})$ is proportional to identity. 
For general $D(\textbf{x})$, the minimization of action function Eq.\eqref{action_Ito}  
does not directly equal to the function $U(\textbf{x})$ even in the limit of $\epsilon\rightarrow0$. For example, when $D(\textbf{x})$ is a diagonal matrix with distinct constant elements,  the action 
$S_{T}[\textbf{x}]|_{I}\geq\sum_{i}\int_{T_{1}}^{T_{2}}dt(\dot{x}_{i}\partial_{x_{i}}U/D_{ii}-l_{i}\partial_{x_{i}}U/D_{ii})\neq\Delta U(\textbf{x})|^{\textbf{s}^{*}}_{\textbf{x}^{*}}$. On the other hand, if we apply the action without the diffusion matrix as in \cite{zhou2012quasi}, $\hat{S}_{T}(\textbf{x})=\sum_{i}\int_{T_{1}}^{T_{2}}dt(\dot{x}_{i}-f_{i})^{2}/2$,
it does not include the effect of $D(\textbf{x})$. A more detailed comparison is given in Sect.~V of Supplemental Information.

Recently, the existence of decomposition Eq.~\eqref{Langevin_decomposed} and its equivalence to Freidlin-Wentzell's quasi-potential in the zero noise limit  is discussed  \cite{zhou2016construction}. Here, we provide a framework to directly extend the potential function to the cases with finite noise. Besides, the classical Freidlin-Wentzell's quasi-potential is locally constructed around each stable attractors, and complicated methods to glue the locally constructed quasi-potentials are required  \cite{freidlin2012random}. In this paper, we find that a unique potential function can be obtained by finding the ideal least action path for the minimization in Eq.~\eqref{minimization}. Even $Q$ in Eq.~\eqref{Langevin_decomposed} is generally not uniquely determined for a given Eq.~\eqref{Langevin0} \cite{zhou2016construction}, certain boundary conditions or physical constrains can be added to specify $Q$ in order to guarantee a unique least action  path for Eq.~\eqref{minimization}.

In order to efficiently find the least action path by the time-reversal adjoint dynamics, here we provide a strategy of minimization called ODE-based-adaptive-time method. First, we simulate ODE, Eq.~\eqref{ODE}, to get a path from ${s}^{*}$ (considering ${s}^{*}$ is a saddle point, add a small perturbation to it as the starting point) to a corresponding stable fixed point ${x}^{*}$. We record the total duration for the trajectory as $T_{{s}^{*}\rightarrow{x}^{*}}^{ODE}$. Since the least action path is time-reversal of the adjoint dynamics, its duration and length should be the same as that of the corresponding trajectory for the original ODE.  Therefore, we choose $T_{2}-T_{1}=T_{{s}^{*}\rightarrow{x}^{*}}^{ODE}$, and put the constrain $inf_{\{T_{2}-T_{1}=T^{ODE}_{{s}^{*}\rightarrow{x}^{*}}\}}$ to the minimization  in Eq.~\eqref{minimization}.  Note that the constrain of trajectory time here could not guarantee to find the ideal least action path, however, it can reduce the sampling space of paths connecting ${x}^{*}$ to ${s}^{*}$. To ensure that the ideal least action path is picked, new types of minimization algorithm considering dynamics by $-Q(\textbf{x})\nabla\phi(\textbf{x})$ need to be developed.

To conclude, we have provided a new approach to study multi-stability and stochastic transitions between stable states for SDE modeling. The potential function can be efficiently calculated corresponding to the Lyapunov function of ODE and the landscape by A-type simulation of SDE with arbitrary noise strength. Our method generates consistent predictions on stochastic processes from both ODE modeling on the average behavior as well as SDE modeling including the effect of noise. It gives probability ratios between stable states in high dimensional systems subject to large noise, such that expensive stochastic simulations can be avoided. The results reveals a new mechanism to control the ratios of cell types by manipulating noise intensity. Our approach should also be practically useful to study role of noise in dynamical modeling for other high dimensional stochastic processes.

\section*{Methods}

We have demonstrated in Sect.~III of Supplemental Information that differences of action functions for Eq.~\eqref{Langevin0} with various stochastic integrations can be neglected when $\epsilon\rightarrow0$. 
Therefore, we can choose action with specific stochastic integration for the convenience of numerical calculations.
Here, we adopt the action with Ito's integration \cite{tang2014summing}:
\begin{align}
\label{action_Ito}
S_{T}[\textbf{x}]|_{I}&=\frac{1}{4}\int_{T_{1}}^{T_{2}}\Big|_{I}dt[\dot{\textbf{x}}-\textbf{f}(\textbf{x})]^{\tau}D^{-1}(\textbf{x})[\dot{\textbf{x}}-\textbf{f}(\textbf{x})].
\end{align}
When $\epsilon\rightarrow0$, its minimization equals to the potential difference $\lim_{\epsilon\rightarrow0}S(\textbf{s}^{*}|\textbf{x}^{*})|_{I}=\lim_{\epsilon\rightarrow0}S(\textbf{s}^{*}|\textbf{x}^{*})|_{A}=\Delta\phi(\textbf{x})|^{\textbf{s}^{*}}_{\textbf{x}^{*}}$ with the least action path satisfying $\dot{\textbf{x}}=D\nabla\phi-Q\nabla\phi$, and deviation between the least action path and that by $\dot{\textbf{x}}=-D\nabla\phi-Q\nabla\phi+\epsilon\Delta\textbf{f}$ for Eq.~\eqref{action_A} disappears as well.

Numerically, we use discretized form of Eq.~\eqref{action_Ito} as the object function to minimize. The discretized scheme adopted corresponds to the stochastic integration \cite{tang2014summing}, such as pre-point scheme is needed for Ito's interpretation. Thus, we minimize the discretized action:
\begin{align}
S_{T}[\textbf{x}]|_{I}&=\frac{1}{4}\sum_{k=1}^{K}\Delta t_{k}\Big[\frac{\textbf{x}^{k}-\textbf{x}^{k-1}}{\Delta t_{k}}-\textbf{f}(\textbf{x}^{k-1})\Big]^{\tau}D^{-1}(\textbf{x}^{k-1})\Big[\frac{\textbf{x}^{k}-\textbf{x}^{k-1}}{\Delta t_{k}}-\textbf{f}(\textbf{x}^{k-1})\Big],
\end{align}
where 
we divide the time intervals to be $T_{1}=t_{1}<\cdots<t_{k}<\cdots<t_{K}=T_{2}$ with $\Delta t_{k}=t_{k}-t_{k-1}$. The trajectory $\textbf{x}(t)$ is divided into $K$ pieces with $\textbf{x}^{k}=\textbf{x}(t_{k})$.
Besides, Eq.~\eqref{action_Ito} is explicitly independent of $\epsilon$, and thus its numerical minimization has implicitly taken the limit $\epsilon\rightarrow0$. We choose straight line connecting the two points as initial path, and use the fmincon function in the toolbox of MATLAB to do minimization, where set $T=-T_{1}=T_{2}$. 

For Eq.~\eqref{Langevin0}, when $Q(\textbf{x})\nabla\phi(\textbf{x})$ is dominate compared to $D(\textbf{x})\nabla\phi(\textbf{x})$, the least action path rotates and is relatively long in length. Then, the number of points $K$ is required to be large enough such that the minimization procedure can find out the long least action path. Besides, we find in the numerical experiment that if $T$ is too large, the least action path may pass an additional saddle point (limit cycle) with higher potential energy before going through the expected saddle point. Special care is needed to choose suitable $K$ and $T$ in these cases, and the ODE-based-adaptive-time method proposed in Sect.~\ref{diss} is a candidate to improve the efficiency. 

There are also numerical methods to adjust the grid points on the least action path,  such as the adaptive minimum action method \cite{zhou2008adaptive}. They can be applied to optimize our current numerical code.

\bibliography{bib}

\noindent LaTeX formats citations and references automatically using the bibliography records in your .bib file, which you can edit via the project menu. Use the cite command for an inline citation, e.g.  \cite{Figueredo:2009dg}.

\section*{Acknowledgements}

We thank Hao Ge, Hong Qian, Nigel Goldenfeld, Xue Lei and Xiaojie Qiu for the critical comments. Discussions with Lei Zhang, Tiejun Li, Jianhua Xing, Joseph Zhou, and Site Li are greatly acknowledged. This work is supported in part by the National 973 Project No.~2010CB529200 and by the Natural Science Foundation of China Projects No.~NSFC61073087 and No.~NSFC91029738.

\section*{Author contributions statement}

Y.T, R.Y, G.W and P.A had the original idea for this work. Y.T and R.Y did the theoretical analysis. R.Y and Y.T conducted the numerical implementation. Y.T, R.Y, G.W and X.Z analyzed the biological examples. All authors discussed the results and contributed to writing the article.

\section*{Additional information}
\subsection**{Supplemental Information Information}
Supplemental Information Information accompanies this paper at [URL will be inserted by publisher].
\subsection**{Competing financial interests}
The authors declare no competing financial interests.

\end{document}